\documentclass{ctr}

\usepackage{ctrfont}
\usepackage{natbib}

\usepackage{url}
\usepackage{amsmath}
\usepackage{amssymb}

\usepackage{graphicx}

\newcommand{\SGS}{\mathrm{{\scriptscriptstyle SGS}}}

\title{Requirements and sensitivity analysis of RANS-free wall-modeled LES}
\shorttitle{Requirements and sensitivity analysis of WMLES}
\author{M. Whitmore, A. Lozano-Dur\'{a}n \and P. Moin}
\shortauthor{Whitmore, Lozano-Dur\'{a}n \& Moin}

\begin{document}


\maketitle



\section{Motivation and objectives} 

The objective of a wall-modeled large-eddy simulation (WMLES) is to accurately predict the engineering quantities of interest of a flow, such as lift or drag forces, without requiring prohibitively expensive near-wall mesh refinement. To achieve this, the wall model must accurately estimate the stress and heat transfer at the wall using variables from the outer LES as inputs. Here, we investigate the sensitivity of the wall model input variables to the modeling choices of the outer LES.

Most existing wall models are based on Reynolds-averaged Navier-Stokes (RANS) formulations, either by directly parametrizing an equilibrium mean velocity profile, or by solving a simplified RANS equation~\citep{BosePark2018}. These RANS-based models require the use of \emph{a priori} empirical coefficients calibrated for a particular set of flows. By explicitly choosing an LES filter, \cite{BoseMoin2014} derived the Robin (slip) boundary condition for wall-modeled LES. The slip boundary condition led to the development of dynamic slip wall models \citep{Lozano2017, Bae2018a}, which are free from equilibrium assumption and tunable coefficients inherent to RANS-based approaches. However, RANS-free wall models have been shown to suffer from sensitivities to LES modeling choices, such as subgrid-scale (SGS) models and numerics. Hence, development of robust dynamic wall models requires understanding and mitigating these sensitivities.

Prior studies have been conducted to address sensitivities of traditional wall models to the input variables, i.e., the wall-parallel velocity. \cite{Kawai2012} argued that the near-wall LES solution is necessarily under-resolved and subsequently found that using inputs from a wall-normal height higher than the first off-wall grid cell led to more accurate stress prediction. \cite{Yang2017} studied the effect of spatial and temporal filtering on the wall model input variables and found that this yielded more accurate wall stress predictions. Nonetheless, these studies are limited to tackle sensitivities that are specific to RANS-based wall models. As such, these studies primarily focus on mean velocity profiles close to the wall as this is the information on which RANS-based models primarily depend. In contrast, dynamic wall models estimate the wall stress using a richer set of information from the outer LES solution, which includes velocity fluctuations and derivatives. The wall model prediction depends on the quality of these input variables. While this is, to an extent, true for both dynamic and RANS-based models, it is particularly important for the former since dynamic wall models rely entirely on the outer LES solution and do not use any {\em a priori} information. It follows then that dynamic wall models require higher fidelity of the LES solution from which the wall model input variables are sampled. This makes dynamic wall models more sensitive than RANS-based models, but also enables higher fidelity in their predictions. The central question addressed in this brief is, how are sensitivities introduced into dynamic wall models and how can they be alleviated?

This brief is organized as follows. The methodology is outlined in Section \ref{methodology}. Results and discussion of sensitivity to the SGS model are presented in Section \ref{SGSsensitivity}. Results and discussion of sensitivity to the numerics and mesh topology are presented in Section \ref{Numsensitivity}. Finally, conclusions are offered in Section \ref{conclusion}.

\section{Methodology} \label{methodology}

\subsection{Numerical experiments}

The sensitivities to modeling choices are assessed through wall-modeled LES calculations of turbulent channels. The channels studied are periodic, having the dimensions $2\pi\delta \times 2\delta \times \pi\delta$, where $\delta$ is the channel half-height. All channels presented in this brief are run at $Re_\tau = 4200$. An exact mean wall stress condition is used that imposes a mean wall stress based on DNS data from \cite{Lozano2014}. The exact mean wall stress formulation with a Neumann boundary condition was used previously by \cite{Lozano2019}. Similarly, the channels are driven by fixing the mean centerline velocity to match the DNS data. This approach was preferred over fixing a constant mass flow rate to avoid discretization uncertainties in computing the mass flow rate at the present coarse resolutions. Three eddy viscosity models are used: the dynamic Smagorinsky model \citep{Germano1991}, the Vreman model \citep{Vreman2004}, and the anisotropic minimum dissipation (AMD) model \citep{Rozema2015}.
\begin{table}
    \centering
    \begin{tabular}{ccccc}
        Grid & $\left(L_1 \times L_2 \times L_3\right)/\delta$ & $\Delta/\delta$ & $N_1 \times N_2 \times N_3$ & $N_{total}$ \\ \hline\hline
        FD Stag. & $2\pi \times 2 \times \pi$ & $0.05$  & $128 \times 40 \times 64$ & $327,680$ \\
        FV HCP   & $2\pi \times 2 \times \pi$ & $0.05$  &                           & $430,591$ \\
        FV Cart. & $2\pi \times 2 \times \pi$ & $0.045$ & $140 \times 44 \times 70$ & $431,200$ \\ \hline
    \end{tabular}
    \caption{LES grid details.}
    \label{tab:les_grids}
\end{table}

\begin{figure}
    \centering
    \includegraphics[width=0.9\textwidth]{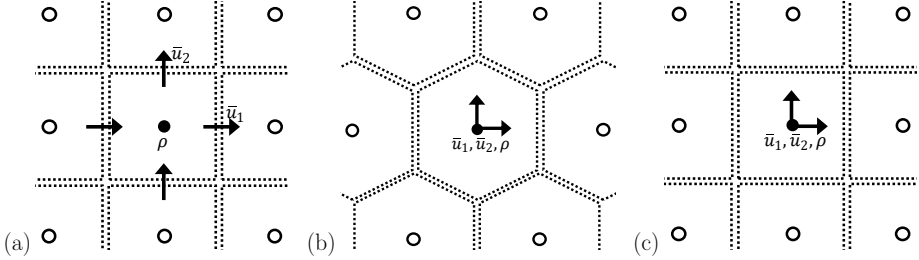}
    \caption{Sketches of slices of the (a) Cartesian staggered grid, (b) HCP grid and (c) Cartesian collocated grid with variable locations indicated.}
    \label{fig:grid_sketch}
\end{figure}

In addition to assessing the effects of SGS model choice, different numerics and mesh topologies are investigated by the use of two different LES codes. The first code is an incompressible flow solver that uses second-order central finite differences with third-order Runge-Kutta time stepping~\citep{Lozano2016}. This code uses a staggered Cartesian grid. The second code is charLES (Cascade Technologies, Inc), which is a fully compressible flow solver using a low-dissipation finite-volume scheme. This code is an unstructured solver that uses Voronoi tessellation capable of generating meshes for arbitrary complex geometries~\citep{Bres2018}. Two point-seeding strategies are used to generate channel meshes: the first is a hexagonal close-packed (HCP) collocated mesh, the standard meshing strategy for charLES, and the second is a basic Cartesian collocated mesh. Details of the LES grids are given in Table \ref{tab:les_grids}. Notional slices of the three meshes are shown in Figure \ref{fig:grid_sketch}(a,b,c).

The staggered Cartesian grid is taken as the baseline grid, with an approximately isotropic resolution of $\Delta \approx 0.05\delta$. The HCP grid was generated using regular point seeding, with a nominal isotropic grid scale of $\Delta=0.05\delta$ to match the staggered Cartesian grid. The Cartesian collocated mesh is chosen such that the total number of control volumes is approximately equal to that of the HCP mesh so as to match the degrees of freedom. HCP grids have higher packing density than a Cartesian grid and the HCP grid would have more mesh points if the nominal grid scales were matched. Instead, the grid scale is chosen to be approximately isotropic with $\Delta\approx0.045\delta$, which results in a similar number of control volumes.

\subsection{Modeling framework}

By applying a low-pass filter operation $\overline{\left(\cdot\right)}$ to the incompressible Navier-Stokes and continuity equations, we obtain the LES governing equations
\begin{equation}
    \frac{\partial \overline{u}_i}{\partial t}
    + \frac{\partial \overline{u}_i\overline{u}_j}{\partial x_j}
    = -\frac{1}{\rho}\frac{\partial \overline{p}}{\partial x_i}
    - \frac{\partial \tau^{\SGS}_{ij}}{\partial x_j}
    + \nu \frac{\partial^2 \overline{u}_i}{\partial x_j \partial x_j} ; \quad
    \frac{\partial \overline{u}_i}{\partial x_i} = 0 ,
\end{equation}
where $\overline{u}_i$ and $\overline{p}$ denote the filtered velocity and pressure fields, $\rho$ is the fluid density, and $\nu$ is the fluid kinematic viscosity. The SGS stress tensor is defined as $ \tau^{\SGS}_{ij} = \overline{u_i u_j} - \overline{u}_i\overline{u}_j $. This term is unclosed and must be modeled. In this investigation, we use eddy viscosity models that relate the anisotropic part of the SGS stress tensor to the resolved rate-of-strain tensor by an eddy viscosity, $\nu_t$. The determination of the eddy viscosity is left to the SGS model.

Two different wall boundary conditions are used in this investigation. The first is the slip boundary condition given by the form
\begin{equation}
    \left.\overline{u}_i\right|_w = \alpha \left.\frac{\partial \overline{u}_i}{\partial n}\right|_w, ~i=1,2,3 ,
    \label{eq:slip_bc}
\end{equation}
where $\left.\left(\cdot\right)\right|_w$ denotes quantities evaluated at the wall and the slip lengths are isotropic, equal to $\alpha$ and constant in space. Previous studies of the slip boundary condition have shown it to have desirable properties such as recovering the no-slip boundary condition in the limit of zero slip length as well as accurately predicting turbulent intensities near the wall \citep{Bae2018b}.

The second boundary condition is the Neumann boundary condition given by the form
\begin{equation}
    \left.\frac{\partial \overline{u}_1}{\partial n}\right|_w = \alpha ,~
    \left.\overline{u}_2\right|_w = 0 ,~
    \left.\frac{\partial \overline{u}_3}{\partial n}\right|_w = \beta ,
    \label{eq:neumann_bc}
\end{equation}
where $\alpha$ and $\beta$ are values used to set the components of the wall-parallel stress in the streamwise and spanwise directions. A no-penetration condition is enforced for the wall-normal velocity. The Neumann boundary condition is present in most widely used wall models and is therefore known to be robust in many cases.

In the context of dynamic wall models, both the slip and the Neumann boundary conditions can be thought of as methods for imposing the wall stress. It is important then to discuss how each boundary condition relates to the wall stress in the mean stress balance. By taking a planar mean in the streamwise and spanwise directions and evaluating at the wall, we obtain the mean stress balance
\begin{equation}
    \langle\tau_w\rangle 
    = \bigg\langle\nu \left.\frac{\partial \overline{u}_1}{\partial n}\right|_w\bigg\rangle 
    - \langle\left.\tau_{1n}^{\SGS}\right|_w\rangle 
    - \langle \left.\overline{u}_1\overline{u}_n\right|_w\rangle ,
    \label{eq:stress_balance}
\end{equation}
where $\langle\cdot\rangle$ denotes the averaging operation, $n$ denotes the wall-normal direction, and $\tau_w$ denotes the total mean wall stress. The first term on the right-hand side represents the viscous stress at the wall, the second term is the SGS stress at the wall, and the third term is the resolved Reynolds stress at the wall. When the Neumann boundary condition is used, the resolved stress term goes to zero due to the no-penetration condition. Thus, for an eddy viscosity SGS model, the mean wall-normal streamwise velocity gradient relates to the total wall stress by a sum of molecular and eddy viscosities. Note that the slip boundary condition interacts with the stress balance primarily by imposing a nonzero resolved stress at the wall. Combining Eq. (\ref{eq:slip_bc}) into the resolved stress term of Eq. (\ref{eq:stress_balance}), we can see that the stress is approximately quadratically related to the isotropic slip length. As a consequence, isotropic slip lengths can only act to increase the total wall stress. This is an important observation because the range of realizable wall stress values is limited to stresses greater than or approximately equal to the sum of the viscous and SGS stresses. More details about this behavior can be found in \cite{Bae2018a}. In theory, the use of two different slip lengths could drain excess stress in a dynamical setting but that method is not explored here \citep{Bae2018a}.

\subsection{Cases and variables for sensitivity analysis}

\begin{table}
    \centering
    \begin{tabular}{ccccccc}
        Case    & SGS model  & Numerics   & Grid topology     & Boundary condition \\ \hline\hline
        S-S-VRE & Vreman     & Fin. diff. & Stagg. Cart.      & Slip    \\
        S-S-DSM & Dyn. Smag. & Fin. diff. & Stagg. Cart.      & Slip    \\
        S-S-AMD & AMD        & Fin. diff. & Stagg. Cart.      & Slip    \\
        N-S-VRE & Vreman     & Fin. diff. & Stagg. Cart.      & Neumann \\
        N-S-DSM & Dyn. Smag. & Fin. diff. & Stagg. Cart.      & Neumann \\
        N-S-AMD & AMD        & Fin. diff. & Stagg. Cart.      & Neumann \\
        N-C-VRE & Vreman     & Fin. vol.  & Colloc. Cart.     & Neumann \\
        N-H-VRE & Vreman     & Fin. vol.  & Hex. close-packed & Neumann \\ \hline
    \end{tabular}
    \caption{LES case details.}
    \label{tab:les_cases}
\end{table}

\begin{figure}
    \centering
    \includegraphics[width=0.8\textwidth]{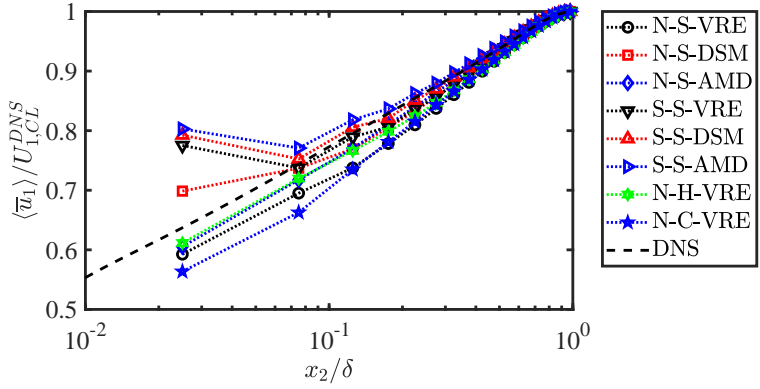}
    \caption{Mean streamwise velocity profiles for all present LES cases.}
    \label{fig:umean}
\end{figure}

We consider 8 WMLES cases. Details of the LES cases are given in Table \ref{tab:les_cases}, which outlines their respective modeling choices. Abbreviated identifiers are given to each case and referenced later. The mean velocity profiles of all WMLES cases are plotted against the DNS mean velocity profile in Figure \ref{fig:umean}. All profiles agree well with the DNS in the outer region, within $\approx \pm 5\%$ error. Closer to the wall at about the fifth off-wall grid point, the profiles become more dispersed, within a range of $\approx +10\%,-15\%$ error, and, notably, the cases with the slip boundary condition have an overshoot in the first off-wall grid cell.

\begin{figure}
    \centering
    \includegraphics[width=0.8\textwidth]{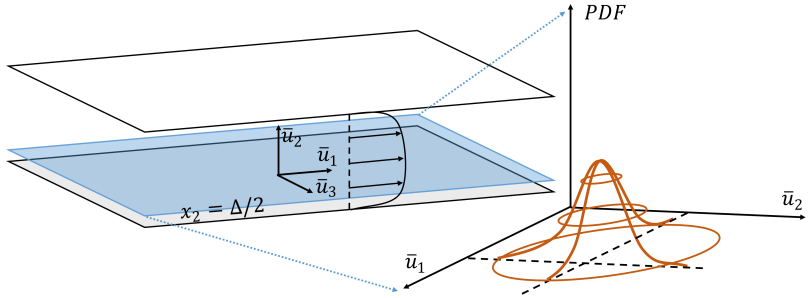}
    \caption{Sketch of the computational domain. The highlighted plane is the region sampled to compute the PDFs.}
    \label{fig:pdf_intro}
\end{figure}

Sensitivities are investigated by inspection of the instantaneous velocity components and the velocity gradients. From the channel simulation data, probability density functions (PDFs) are computed for LES velocities and their gradients. A sketch of the setup is shown in Figure \ref{fig:pdf_intro}. To consolidate the velocity gradient information, PDFs of the second and third invariants of the velocity gradient tensor, $Q$ and $R$, are presented. All quantities are normalized by outer flow scales, with the local mean velocity $U_1^{DNS}$ and the mean centerline velocity $U^{DNS}_{1,CL}$ taken from the DNS data. This normalization allows the PDFs to show both the fluctuations and the qualitative accuracy of the LES velocities in the mean. The data used to compute the PDFs are taken from a plane parallel to the wall at some wall-normal height. Since we are interested in observing sensitivities in the context of dynamic wall models, the planes near the wall are of the most interest because these are the data on which the wall model relies to make predictions of wall quantities.

We focus mainly on data sampled from the height of the first off-wall grid cell. Considering that one objective of a dynamic wall model is to simulate complex geometries, the first off-wall grid cell is the ideal sampling location because finding the second or third off-wall grid cell is not necessarily a well-defined problem for arbitrary geometries. Incidentally, the planes near the wall are expected to show the highest sensitivity as numerical errors as well as SGS modeling errors are larger near the wall.

\section{Sensitivity to SGS model} \label{SGSsensitivity}

\subsection{SGS sensitivity with a Neumann boundary condition}

\begin{figure}
    \centering
    \includegraphics[width=0.9\textwidth]{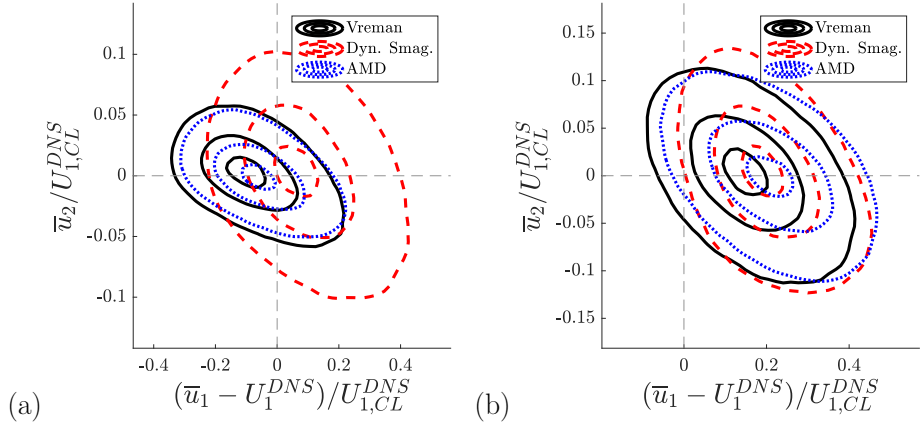}
    \caption{Joint PDFs of normalized streamwise and wall-normal velocities for the $Re_\tau=4200$ 
    channel at the (a) first ($x_2=0.5\Delta$) and (b) third ($x_2=2.5\Delta$) 
    grid points. Vreman, dynamic Smagorinsky, and AMD models are used with a 
    Neumann boundary condition.}
    \label{fig:matching_location}
\end{figure}

\begin{figure}
    \centering
    \includegraphics[width=0.7\textwidth]{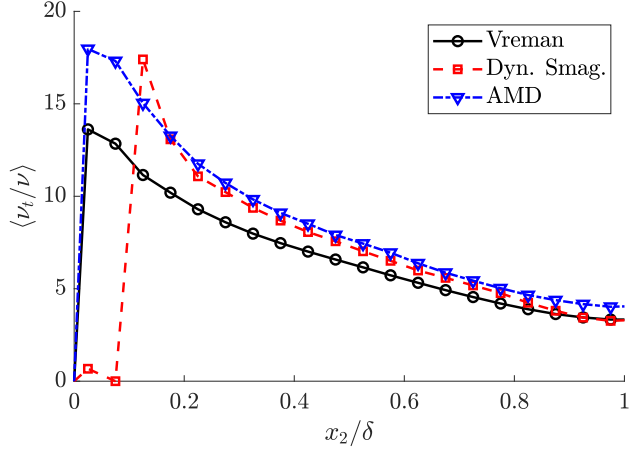}
    \caption{Mean eddy viscosity as a function of wall-normal distance for cases N-S-VRE, N-S-DSM and N-S-AMD. Normalized by molecular viscosity.}
    \label{fig:neumann_eddy_visc}
\end{figure}

Turbulent channels are simulated using the three different SGS models on the staggered grid with the wall stress imposed by a Neumann boundary condition. Contours of the joint PDF of streamwise and wall-normal velocities appear in Figure \ref{fig:matching_location}(a), comparing cases N-S-VRE, N-S-DSM and N-S-AMD. The contours in the velocity PDF plots represent $\left[0.9,0.5,0.1\right]\times$ the maximum value of the PDF. There is a strong difference in the computed PDFs, especially comparing the dynamic Smagorinsky result with the Vreman and AMD results. The magnitude of velocity fluctuations is much larger for the dynamic Smagorinsky, particularly in the wall-normal component, which results in a different slope of the PDF. One explanation for this result is due to the near-wall behavior of the SGS models. The eddy viscosities predicted in each of these cases are shown in Figure \ref{fig:neumann_eddy_visc}. The eddy viscosity predicted by the dynamic Smagorinsky model decays to a lower value in the cells near the wall, than do the eddy viscosities predicted by the Vreman and AMD models, which reach a peak in the first off-wall cell. The net effect is that the velocity fluctuations of the latter models are damped similarly near the wall while the dynamic Smagorinsky model maintains larger velocity fluctuations.

PDFs of the channel solutions at different wall-normal heights are shown in Figure \ref{fig:matching_location}(a,b). As expected, the joint PDFs of streamwise and wall-normal velocities at the height of the first off-wall grid cell show strong sensitivity to the SGS model. However, the joint PDFs at the height of the third off-wall grid point have an improved collapse. This result demonstrates that at increasing wall-normal height, the sensitivity to the SGS model diminishes as the grid is able to resolve more of the energy-containing scales of the flow. This is consistent with previous results in the context of the equilibrium wall model: sampling data from the second or third off-wall grid point leads to better prediction of the wall stress since the flow near the wall is necessarily under-resolved \citep{Kawai2012,Larsson2016}. This result is expected to apply generally, that sampling from a higher wall-normal location will give data that are less contaminated with numerical and modeling errors. This approach does have a drawback, though, in the case of nonequilbrium flows. For example, for a flow with a 3D mean profile it would be better to use information from the first off-wall grid cell to better predict the shear stress direction.

Experiments into the use of temporal filtering in the course of this investigation found that its effect was anisotropic, acting dominantly in the streamwise direction, and was not sufficient to mitigate the sensitivities observed. These experiments were motivated by the previous study on filtering by \cite{Yang2017}, but the present results did not show it to be as effective. Investigations into wall-parallel spatial filtering found small mitigation of sensitivities. This approach, however, does not extend well to complex geometries and, thus, is not expected to be a general solution.

\subsection{SGS sensitivity with a slip boundary condition}

\begin{figure}
    \centering
    \includegraphics[width=0.9\textwidth]{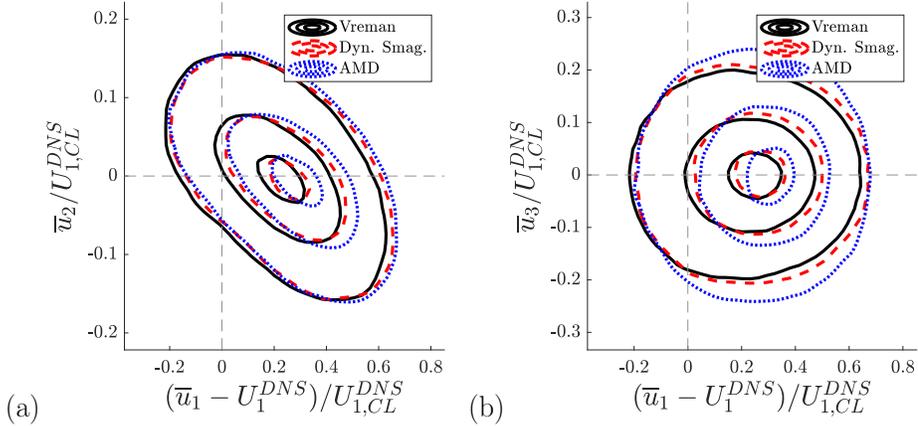}
    \caption{Joint PDFs of (a) streamwise and wall-normal velocities and (b) streamwise and spanwise velocities at the height of the first off-wall grid cell using three different SGS models. A slip boundary condition is used, with $f=0$. }
    \label{fig:slip_pdfs}
\end{figure}

Three calculations with different SGS models are run using a slip boundary condition to impose the wall stress. Again, the staggered grid is used. The cases correspond to S-S-VRE, S-S-DSM and S-S-AMD and the PDFs are computed using data from the first off-wall grid cell. In the first attempt at running these cases, it was found that only the case with dynamic Smagorinsky could impose the desired wall stress. The other two cases failed to impose the desired wall stress because the SGS stress at the wall was too large. In that sense, these cases using a slip boundary condition showed sensitivity to the SGS model.

To understand the sensitivity to SGS model, let us consider Eq. (\ref{eq:stress_balance}), and assume that the total wall stress is fixed around a desired value. Noting that the viscous stress term is small relative to the other terms, it is apparent that the resolved Reynolds stress term will be sensitive to changes in the SGS model because it will respond to changes in the SGS stress at the wall. For this reason we define a parameter
\begin{equation}
    f \equiv \frac{-\langle \left.\tau_{1n}^{SGS}\right|_w \rangle}{\langle \tau_w \rangle},
    \label{eq:f_def}
\end{equation}
where $f$ denotes the fraction of the total wall stress that is carried by the SGS model. This parameter plays an important role in affecting the near-wall resolved scales, because it effectively constrains the amount of resolved Reynolds stress at the wall, which then interacts with the near-wall flow and can introduce fluctuations.

\begin{figure}
    \centering
    \includegraphics[width=0.9\textwidth]{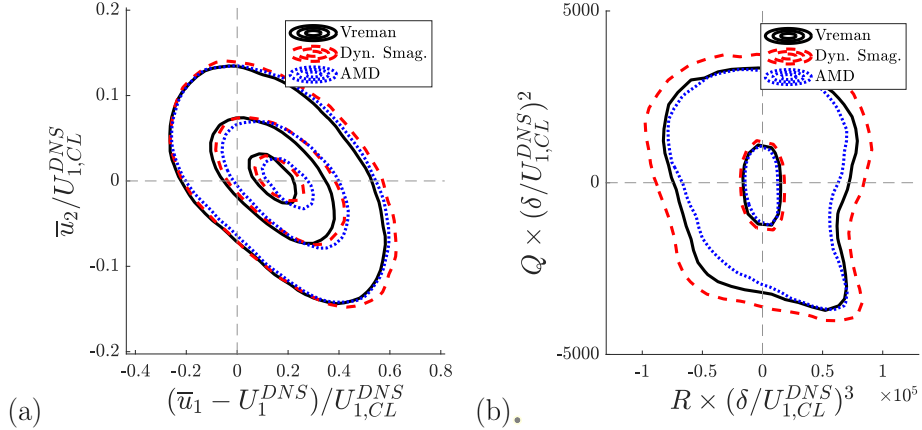}
    \caption{Joint PDFs of (a) streamwise and wall-normal velocities and (b) second and third invariants of the velocity gradient tensor. Slip boundary condition imposes wall stress, the SGS model is varied, and $f\approx0.3$.}
    \label{fig:slip_pdfs_f03}
\end{figure}

Fixing the parameter $f=0$, the wall model is able to impose the desired wall stress for all three cases. Joint PDFs of the streamwise and wall-normal velocities, as well as the streamwise and spanwise velocities, are shown in Figure \ref{fig:slip_pdfs}(a,b). The result in these plots, in comparison to that for the Neumann boundary condition cases, shows much less sensitivity to changing the SGS model. 

The collapse in these results is further observed for values of $0 \leq f \lesssim 0.5$ so long as $f$ is held constant between the calculations. Results shown in Figure \ref{fig:slip_pdfs_f03}(a,b) demonstrate for $f\approx0.3$ a qualitatively very good collapse in the joint PDFs of streamwise and wall-normal velocities and in the joint PDFs of the second and third invariants of the velocity gradient tensor, where contours represent $[0.5, 0.05]\times$ the maximum value of the PDF. For the values tested, $f\approx \left[0,0.3,0.5,0.8\right]$, the collapse starts to deteriorate around $f\approx0.8$. This is due to the effect that, as $f$ increases, the stress at the wall is carried increasingly by the SGS model and the solution at the first off-wall grid point becomes more strongly influenced by the SGS model than by the resolved stress at the wall.

\section{Sensitivity to numerics and mesh topology} \label{Numsensitivity}

\begin{figure}
    \centering
    \includegraphics[width=0.9\textwidth]{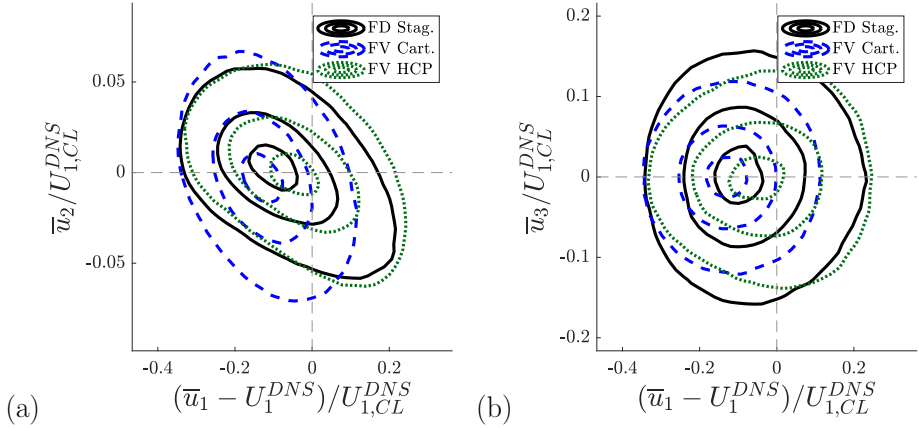}
    \caption{Joint PDFs of (a) streamwise and wall-normal velocities and (b) streamwise and spanwise velocities. The Vreman model is used with a Neumann boundary condition.}
    \label{fig:numerics_pdfs}
\end{figure}

Turbulent channels are run with two different LES codes in order to assess sensitivity to the numerical scheme. In addition, three different grid topologies, the staggered Cartesian grid in the finite-difference code and the collocated HCP and collocated Cartesian grids in the finite-volume code, are tested. Figure \ref{fig:numerics_pdfs}(a,b) shows the joint PDFs of LES velocities compared across cases N-S-VRE, N-C-VRE and N-H-VRE. Taking the staggered Cartesian grid case to be the baseline, sensitivity is observed when changing to the finite-volume code using the collocated Cartesian mesh. The PDF from the collocated Cartesian mesh has a different slope from that for the staggered grid. This implies an excess of strong wall-normal velocity fluctuations with a deficit of streamwise velocity fluctuations. A deficit of spanwise velocity fluctuations is also observed when comparing the collocated Cartesian grid to the staggered Cartesian grid. In contrast, the HCP grid gives results qualitatively similar to those of the staggered grid. This is seen particularly in the joint PDFs of streamwise and wall-normal velocities where the PDF contours exhibit roughly the same shape and magnitude. In the spanwise fluctuations, the HCP grid has a deficit relative to that of the staggered grid, but not as severe as that of the collocated Cartesian mesh. In the context of these results it is apparent that whereas the Cartesian collocated grid does show sensitivity, the HCP grid does not show a strong sensitivity in the LES velocities relative to the staggered grid results.

In comparing the Cartesian collocated mesh to the HCP mesh, it is important to note that although the total number of control volumes in each mesh is approximately the same, the meshes still differ significantly in the number of flux face areas. While the Cartesian collocated cells are cubes having only 6 faces, the HCP cells are truncated octahedra that have a total of 14 faces. The implication of this is that the HCP mesh has more than twice as many flux calculations in total and that this allows the HCP mesh to resolve more flow directions compared to the collocated Cartesian mesh. For the present simulations, in which the computational cost is bounded by the flux calculations, this results in an approximately $2\times$ increase in computation time. However, for a calculation that is bounded by the wall model computations, the change in computational cost would be negligible as the number of boundary faces does not change significantly between mesh topologies.

\begin{figure}
    \centering
    \includegraphics[width=0.9\textwidth]{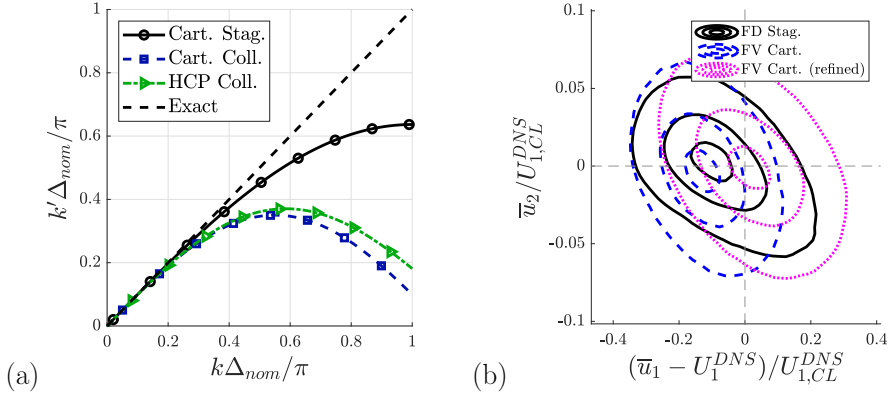}
    \caption{(a) Modified wavenumber $k^\prime$ versus wavenumber $k$ evaluated analytically for three different mesh topologies, normalized by the nominal grid scale $\Delta_{nom}$ equal to that used by the staggered grid. (b) Joint PDFs of streamwise and wall-normal velocities showing the effect of $\sim 2\times$ global grid refinement in the Cartesian collocated grid case. The Vreman model is used with a Neumann boundary condition and the data are sampled at the height of the first off-wall grid cell of the coarser mesh.}
    \label{fig:refined_cart}
\end{figure}

An additional consideration in comparing the collocated Cartesian grid to the staggered grid is the idea of effective resolution. A simplified analytical argument about effective resolution can be made in terms of the modified wavenumbers of the discrete derivative operator. In Figure \ref{fig:refined_cart}(a), the modified wavenumbers of a second-order central difference operator are plotted for the three present meshes. This plot suggests that the collocated Cartesian mesh would need to have roughly twice the resolution of the staggered Cartesian mesh in order to resolve the same amount of wavenumber content. This argument does not apply exactly because it does not accurately represent the operators used in the finite volume code, however, we can assess the relevance of the effective resolution by applying a $\sim 2\times$ global refinement to the collocated Cartesian mesh and running the case with the finite volume code. In Figure \ref{fig:refined_cart}(b), joint PDFs are plotted showing the change when the collocated Cartesian grid is refined. While the slope of the PDF for the refined collocated Cartesian case has gotten closer to that of the staggered mesh, it still has not collapsed and the shapes of the PDFs are qualitatively different. This suggests that while the effective resolution of the mesh plays a small role, it is not the primary effect introducing the sensitivity. For this reason it is suspected that the number of flux face areas is the primary effect driving the sensitivity.

\section{Conclusions} \label{conclusion}

Development of robust dynamic wall models requires understanding the sensitivities of the near-wall WMLES solution to modeling choices. This is because dynamic wall models rely on these flow quantities as input variables. In order to assess these sensitivities to modeling choices, we used turbulent channel calculations. The turbulent channel calculations were set to impose the correct wall stress and correct centerline velocity by taking values from DNS data. Multiple channels were run in this configuration, perturbing modeling choices including SGS model, boundary condition type, numerics and mesh topology. PDFs were computed using outer LES information from wall-parallel planes at some wall-normal sampling height, typically the height of the first off-wall grid cell.

When testing three different SGS models with wall stress imposed by a Neumann boundary condition, we observed sensitivity. This sensitivity was mitigated by increasing the wall-normal distance of the sampling location for the input to the wall model. This result is in agreement with previous results for traditional wall stress models due to the near-wall flow being necessarily under-resolved in the first off-wall grid cell \citep{Kawai2012,Larsson2016}. However, this method of raising the wall-normal sampling location has the drawback that, for nonequilibrium flows, it is not as accurate as using the first off-wall grid cell as input to the wall model. 

Using a slip boundary condition and perturbing the SGS model choice, we observed that the sensitivity can be mitigated even in the first off-wall grid cell. This was achieved by fixing the fraction $f$ of the wall stress carried by the SGS model across calculations. Simulations showed that fixing $f$ in the range $0\leq f\lesssim 0.5$ effectively mitigated the sensitivity to the SGS model. Ultimately, use of the slip boundary condition is desirable for development of robust dynamic wall models since it allows mitigation of sensitivities in the first off-wall grid cell which is helpful for extension to nonequilibrium flows and complex geometries. An additional approach not studied here connects the Neumann boundary condition case to results from the slip boundary condition case. When using the Neumann boundary condition, modifying the eddy viscosity value on the wall can change the slope of the mean velocity profile at the wall. The dynamical response of changing the slope could be significant because of the impact on turbulent production. Evaluating this effect and its role in sensitivities is a topic for future investigation.

Three calculations were run to test the sensitivity of the LES solution to numerics and mesh topology. Sensitivity was observed when changing from the staggered Cartesian mesh to the collocated Cartesian mesh. However, no sensitivity was observed when changing from the staggered Cartesian mesh to the collocated HCP mesh. This result was found even though the collocated Cartesian mesh and the collocated HCP mesh were chosen to have approximately the same number of degrees of freedom.

\section*{Acknowledgments} 

This investigation was funded by the Stanford Engineering Graduate Fellowship and by NASA grant \#NNX15AU93A. 

\bibliographystyle{ctr}


\end{document}